\documentclass[conference]{IEEEtran}
\usepackage{cite}

\usepackage{amsfonts}

\usepackage{graphicx}

\usepackage[cmex10]{amsmath}

\usepackage{amssymb, bm, amsthm}

\usepackage{algorithmic}

\usepackage{algorithm}

\usepackage{color}

\usepackage{mathrsfs}

\ifCLASSINFOpdf
\else
\fi

\begin{document}
%
\title{A TDMA-like Access Scheme with Splitting Request and Transmission for Vehicular Networks}

\author {
        \IEEEauthorblockN{Shenglong Peng$^\dagger$, Xuan He$^\dagger$, Junyi Du$^\ddagger$, Yong Liang Guan$^\S$, and Liang Zhou$^\dagger$}
        \IEEEauthorblockA{$^\dagger$National Key Lab of Science and Technology on Communications, Center for Intelligent Networking and Communications,\\
        Center for Cyber Security, University of Electronic Science and Technology of China, Chengdu, China\\
        $^\ddagger$ CETC Key Lab of Avionic Information System, Southwest China Institute of Electronic Technology, Chengdu, China\\
        $^\S$ School of Electrical and Electronic Engineering, Nanyang Technological University, Singapore}
        Emails: psl880307@163.com, helaoxuan@126.com, jydu1989@163.com, eylguan@ntu.edu.sg, lzhou@uestc.edu.cn
        }

\maketitle

\begin{abstract}
In this paper, we consider safety message transmission in a dense vehicular network. With increasing vehicular network density, the collision rate increases when multiple vehicles transmit safety messages simultaneously. To address this issue, we propose a request-transmission split time division multiple access (TDMA) scheme, referred to as RTS-TDMA. In our scheme, we divide a frame into three phases, i.e., a contention access phase, a broadcast feedback phase, and a contention-free transmission phase. Each vehicle selects a repetition rate according to a given probability distribution and repeats the transmission of its request packet to improve the reliability of the request. In addition, a roadside unit acts as the coordinator and uses a successive interference cancellation technique to resolve request collisions. RTS-TDMA also reduces the request time percentage by containing only the vehicle identity in each request packet. Both theoretical analysis and numerical results verify that the RTS-TDMA scheme can provide higher throughput than the coded slotted ALOHA scheme.\\
\end{abstract}


\IEEEpeerreviewmaketitle

\section{Introduction}
A vehicular network is a special network in which vehicles can transmit data to each other via vehicle-to-vehicle (V2V) communication and to the infrastructure via vehicle-to-infrastructure (V2I) communication \cite{zhang2011data}. Based on these two types of communication, many real-world applications can be supported in vehicular networks. According to their properties, the applications can be classified into two categories: safety (such as accident prevention) and nonsafety (such as Internet access). Dedicated short range communication (DSRC), which is exclusively used by V2V and V2I communications, has been allocated 75 MHz of radio spectrum in the 5.9 GHz band. The spectrum is divided into seven channels, i.e., one control channel (CCH) and six service channels (SCHs)\cite{Hadded2015survey}. The CCH is used to transmit control messages or safety messages, while the SCHs are used to transmit nonsafety messages. Compared to nonsafety messages, safety messages have stricter latency and reliability requirements; e.g., the lifetime of a safety message is less than 100 ms \cite{camp2005vehicle}. Thus, designing a practical scheme for disseminating safety messages is an urgent need and has attracted considerable attention\cite{IEEE802.11p,Eichler2007,Zheng2016,borgonovo2004adhoc,Omar2013vemac,Lyu2018}.

According to the adopted medium access control (MAC) protocols, existing works can be divided into two categories: contention-based MAC protocols \cite{IEEE802.11p,Eichler2007,Zheng2016} and contention-free MAC protocols \cite{borgonovo2004adhoc,Omar2013vemac,Lyu2018}. In contention-based MAC protocols, each vehicle employs a carrier-sense multiple access with collision avoidance (CSMA/CA) mechanism to avoid access collisions among vehicles. When a vehicle attempts to access the medium, it first listens to the channel. If the channel is idle, the vehicle can access the medium; otherwise, the vehicle has to randomly select a back-off time until the channel is idle. In addition, by using the enhanced distributed channel access (EDCA) scheme in IEEE 802.11p, a high-priority safety message is allocated to the high-priority access category (AC). Thus, a vehicle that has a high-priority safety message can use a small contention window size to compete for the channel. In contention-free MAC protocols, each vehicle is required to broadcast the usage status of time slots used by its one-hop neighbors. Thus, each vehicle can access an idle time slot that is not occupied by its two-hop neighbors in a distributed manner. If successful, it will continue to access the same time slot in all subsequent frames until a collision occurs.

However, existing protocols may not work well in a congested scenario in which safety messages from multiple vehicles collide. For contention-based MAC protocols, with increasing vehicular network density, multiple vehicles may transmit safety messages simultaneously even with a back-off mechanism and a small contention window size. As a result, unbounded delays in the dissemination of safety messages occur in this type of vehicular network\cite{Eichler2007}. For contention-free MAC protocols, with increasing vehicular network density, the probability of multiple vehicles choosing the same idle time slot will increase significantly. As a result, safety messages may fail to disseminate in this type of vehicular network\cite{Omar2013vemac}.

Recently, coded slotted ALOHA (CSA) has been widely used in wireless communication systems to solve user collision issues and serve a large number of users \cite{Paolini2015CSA}. Before transmission, the packet from each user is divided and encoded into multiple packets via local component codes at the MAC layer. By combining the successive interference cancellation (SIC) technique and the decoding of packet-oriented codes on the receiver side, collided packets can be recovered\cite{Sun2017Design}. Inspired by the CSA scheme, we introduce the SIC technique to vehicular networks and propose a \underline{r}equest-\underline{t}ransmission \underline{s}plit time division multiple access (\underline{TDMA}) scheme, referred to as \textit{RTS-TDMA}. Note that our proposed RTS-TDMA scheme differs from CSA as well as the other existing MAC schemes. We list the differences and our contributions as follows:

\begin{itemize}
    \item We divide a frame into three phases, i.e., a contention access phase, a broadcast feedback phase, and a contention-free transmission phase.
    \item Each vehicle selects a repetition rate according to a given probability distribution and repeats the transmission of its request packet. Each packet contains only the vehicle identity (ID) and the location information of the contention resources occupied by other copies, thus allowing vehicles to achieve time resource reductions during requests.
    \item Roadside units (RSUs) employ the SIC technique to eliminate collisions among request packets and extract vehicle IDs. Then, each RSU assigns time slots for vehicles according to the extracted ID order.
    \item Both theoretical analysis and simulation results show that the proposed RTS-TDMA scheme provides significantly higher throughput than CSA. 
\end{itemize}

The remainder of this paper is organized as follows. Section II describes the system model. Section III presents our proposed design approach. Section IV provides the performance analysis, and conclusions are drawn in Section V.\par

\section{System Model}

\begin{figure}
\centering
\includegraphics[width=7.6cm,height=4cm]{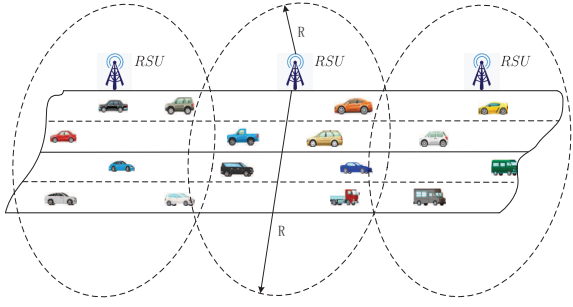}
\caption{Vehicular network scenario}
\label{fig: VANET}
\end{figure}

We consider a vehicular network scenario, as shown in Fig. \ref{fig: VANET}, where $V$ vehicles are moving on a highway. Each vehicle has its own unique ID, where $ID \in \{ 1,2,\dots,V \}$. $U$ RSUs divide a road into $U$ different \textit{service zones} (SZs). Each RSU is used as a central coordinator to serve the vehicles in an SZ of length $2 \times R$ meters, where $R$ is the communication range. Note that we consider only the communication process in a single SZ and try to improve its throughput performance. The overlap between two neighboring RSUs is not taken into consideration in this paper, but we plan to consider it in our future work. \par


\begin{figure}
\centering
\includegraphics[width=8.15cm,height=2.1cm]{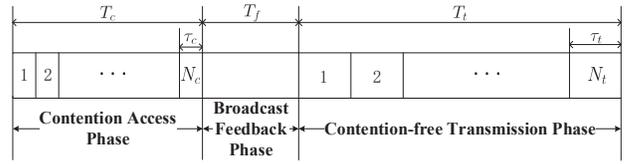}
\caption{The structure of one frame}
\label{fig: frame}
\end{figure}

 Time is partitioned into frames. Each frame is further partitioned into three phases: 1) the contention access phase (\textit{CAP}), 2) the broadcast feedback phase (\textit{BFP}), and 3) the contention-free transmission phase (\textit{CTP}). The durations of these phases are denoted by $T_{c}$, $T_{f}$, and $T_{t}$, respectively, as shown in Fig. \ref{fig: frame}. The CAP is divided into $N_c$ equally spaced \textit{mini-slots} of duration $\tau_c$. The CTP consists of $N_t$ equally spaced \textit{time slots} of duration $\tau_t$. Each vehicle must acquire exactly one time slot to transmit its safety message within each frame. The vehicles and RSUs are equipped with a global positioning system, which is used for frame and slot synchronization\cite{Omar2013vemac}.\par

 The communication is conducted as follows:
 \begin{itemize}
 \item[1.] In the CAP, each vehicle selects a repetition rate according to a given probability distribution and repeats the transmission of its request packet to an RSU to request a time slot in the CTP. Each packet contains only the vehicle ID and the location information of the mini-slots occupied by the other copies.
 \item[2.] In the BFP, the RSU first uses the SIC technique to eliminate collisions among request packets and extracts the IDs of the vehicles. Then, the RSU allocates time slots to the vehicles and broadcasts a feedback packet that consists of a list of vehicle IDs. The maximum number of allowed vehicles is equal to $N_t$, i.e., the number of time slots in the CTP.
 \item[3.] In the CTP, an assigned vehicle occupies its corresponding time slot and disseminates its safety message to the RSU and neighbor vehicles.
 \end{itemize}

\section{Design of the access scheme}
In this section, we describe the details of how vehicles acquire time slots in each CTP with the help of the RSUs and complete the dissemination of their safety messages within each frame.

\subsection{The contention access process in the CAP}
Our proposed RTS-TDMA scheme uses CSA as the contention access scheme in the CAP. However, in RST-TDMA, a request packet contains only an ID and the location information of the mini-slots occupied by its copies. The request packet is much shorter than a packet containing a payload. Therefore, the request packet in RTS-TDMA does not need to be divided. In addition, the request packet in RTS-TDMA is repeated instead of being encoded before transmission, and SIC can be executed once the receiver confirms that no collision has occurred in a time slot. Hence, RTS-TDMA employs repetition codes as the local components of the CSA scheme.

At the beginning of each CAP, a vehicle, for example, $x$, maps a message containing only $ID_x$ to a request packet $P_r(x)$ of the physical layer and draws an ($l$, 1) repetition code, where $l$ is randomly selected based on a probability distribution $\{\Lambda_l\}$. Then, vehicle $x$ copies $P_r(x)$ into $l$ packets, denoted by $\mathbf{P}_r(x)=(P_r^1(x),P_r^2(x),\dots,P_r^l(x))$. Vehicle $x$ transmits to the RSU in the SZ over $l$ mini-slots, which are selected randomly and uniformly from $(1,2,\dots,N_c)$, denoted by $\mathbf{CS}(x)=(i_1,i_2,\dots,i_l)$, where $1\leq i_1 < i_2 < \dots < i_l \leq N_c$. Each packet contains the location information of the mini-slots occupied by other copies, which is called the pointer. We assume that collisions are always detected by RSUs and that the SZ is within the range of detectability and decodability of an RSU. Hence, once the RSU successfully decodes a \textit{clean} packet (i.e., a packet that did not collide), it extracts the pointer of the copies and $ID_x$. Interference cancellation is assumed to be ideal. Suppose that the copy has collided; then, the RSU allows the copy to be subtracted from the received signal in the corresponding mini-slot. Therefore, another packet may be decoded in the same mini-slot. SIC is iterated until all packets have been decoded successfully or until no clean packets exist. All extracted IDs will be added to a slot assignment list by the RSU, denoted by $\mathbf{A}$.\par

\begin{figure}
\centering
\includegraphics[width=7.8cm,height=12.5cm]{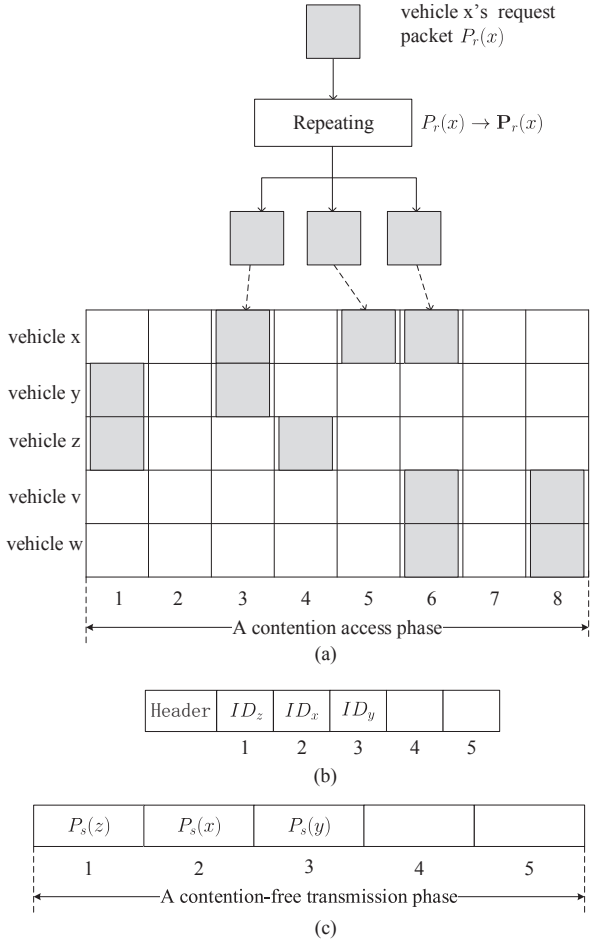}
\caption{An example of the scheme: (a) the process in the CAP; (b) the format of the feedback packet; (c) the assignment of time slots in the CTP.}
\label{fig: example}
\end{figure}

 Fig. \ref{fig: example}(a) shows an example of the process describing competition among vehicles in the CAP. There are $M=5$ vehicles (indexed as vehicle $x$, vehicle $y$, vehicle $z$, vehicle $v$, and vehicle $w$) in the SZ of an RSU. There are $N_c=8$ mini-slots (indexed from 1 to 8) in a CAP and $N_t=5$ (indexed from 1 to 5) in a CTP. All five vehicles are in the coverage area of the RSU. Of the five vehicles, vehicle $x$ employs a (3, 1) repetition code, while vehicle $y$, vehicle $z$, vehicle $v$, and vehicle $w$ employ (2, 1) repetition codes. Then, the packets in $\mathbf{P}_r(x)$ are transmitted in mini-slots \{3, 5, 6\} of the CAP. Similarly, the two packets in $\mathbf{P}_r(y)$ are transmitted in mini-slots \{1, 3\} of the CAP. The two packets of $\mathbf{P}_r(z)$ are transmitted in mini-slots \{1, 4\} of the CAP. The two packets of $\mathbf{P}_r(v)$ are transmitted in mini-slots \{6, 8\} of the CAP. The mini-slots occupied by packets in $\mathbf{P}_r(w)$ are the same as those occupied by packets in $\mathbf{P}_r(v)$. \par

 Decoding is performed on the RSU side as follows. At iteration 1 of SIC, the RSU receives and decodes $P_r^2(z)$ in mini-slot 4. Then, $ID_z$ can be extracted. The RSU knows that $P_r^1(z)$ is transmitted in mini-slot 1 according to the information of the pointer included in $P_r^2(z)$. The contribution of the interference of $P_r^1(z)$ in mini-slot 1 can be subtracted. Similarly, $ID_x$ can be extracted from $P_r^2(x)$ in mini-slot 5. The contributions of the interference of $P_r^1(x)$ in mini-slot 3 and that of $P_r^3(x)$ in mini-slot 6 can be subtracted. The RSU adds vehicle $z$ and vehicle $x$ to $\mathbf{A}$, where $\mathbf{A}=(ID_z, ID_x)$. At iteration 2, both packets of $\mathbf{P}_r(y)$ can be decoded. The RSU extracts $ID_y$ from any packet of $\mathbf{P}_r(y)$. The RSU adds $ID_y$ to $\mathbf{A}$, where $\mathbf{A}=(ID_z, ID_x, ID_y)$. However, any packet that has been received or revealed cannot be used to recover the packets of vehicle $v$ and vehicle $w$. Therefore, the RSU cannot extract $ID_v$ or $ID_w$.\par

\subsection{The broadcast feedback process in the BFP}
At the end of each CAP, all IDs extracted successfully by the RSU have been added to $\mathbf{A}$, where the IDs are listed in the order of when they were extracted. Then, the RSU assigns the $k$th time slot of the CTP to the vehicle with the $k$th ID in $\mathbf{A}$, where $1 \leq k \leq \left|\mathbf{A}\right|$. However, when $N_t < \left|\mathbf{A}\right|$, because the CTP contains $N_t$ time slots, only vehicles with the first $N_t$ IDs in $\mathbf{A}$ will be assigned a time slot of the CTP. Finally, in the BFP, the RSU will broadcast the feedback packet, denoted by $P_{fb}$, to the vehicles in its SZ. In the feedback packet, each field containing the ID of a vehicle represents the assignment of the vehicle to the corresponding time slot in the CTP. \par

The example in Fig. \ref{fig: example} shows that the RSU holds $\mathbf{A}=(ID_z, ID_x, ID_y)$. According to the order of IDs in $\mathbf{A}$, the RSU assigns the time slots of indexes 1, 2 and 3 in the CTP for vehicle $z$, vehicle $x$, and vehicle $y$, respectively. Then, the RSU broadcasts the feedback packet $P_{fb}$ to the vehicles in its SZ, including the time slot assignment information. The format of the feedback packet $P_{fb}$ is shown in Fig. \ref{fig: example}(b). \par

\subsection{The contention-free transmission in the CTP}
When a vehicle, for example, $x$, receives a feedback packet $P_{fb}$ from the RSU, $x$ will check whether it has been assigned a time slot in the CTP. If so, $x$ will disseminate its safety packet $P_s(x)$ during the corresponding time slot. Otherwise, $x$ knows that its ID has not been extracted by the RSU in the previous CAP and that it cannot send messages in the CTP; thus, it must wait for the next frame to request again.\par

The example in Fig. \ref{fig: example} shows that all the vehicles in the SZ of the RSU check the time slots that they have been assigned. As shown in Fig. \ref{fig: example}(c), vehicle $x$, vehicle $y$ and vehicle $z$ then disseminate their safety packets in the time slots of indexes 1, 2 and 3, respectively. However, vehicle $v$ and vehicle $w$ do not check the time slots that they have been assigned, and they cannot transmit their safety messages in the current CTP.\par

\section{Performance analysis}
The objective of the analysis in this section is to determine the value range of the maximum number of vehicles allowed in an SZ and the duration of a frame such that the throughput performance of our scheme will be higher than that of CSA.\par

\subsection{Theoretical analysis}
Above all, we need to determine which parameters are constant and which are variable. We can determine the length of a safety message following \cite{camp2005vehicle}. The transmission rate is provided by the IEEE 802.11p\cite{IEEE802.11p}. Thus, the duration ($\tau_t$) of a time slot in the CTP is constant. In addition to the pointer, only a vehicle ID is transmitted in the mini-slots of each CAP. The duration ($\tau_c$) of a mini-slot in the CAP is also constant. The maximum number of vehicles allowed in an SZ is denoted by $M_{max}$. The number of vehicles in an SZ during a frame is denoted by $M$, where $0 \leq M \leq  M_{max}$. We denote $T_F$ as the duration of a frame. In the analysis, $M_{max}$ and $T_F$ are variables.

The number of slots in each frame of CSA is $N_I=T_F/\tau_I$, where $\tau_I$ is the duration of a time slot in CSA. Because the same safety message is considered to be transmitted, we have $\tau_I=\tau_t$. The throughput $S$ represents the average number of successful safety packet transmissions per frame in an SZ. According to the result in CSA, if $M$ is close to or greater than $N_I$, CSA suffers a significantly poor throughput performance. Therefore, we set the value of $T_F$ such that $T_F > M_{max}\tau_I$. Note that in the analysis, we set the same probability distribution of the repetition rate $l$ in CSA and in the CAP of RTS-TDMA. Thus, the throughput in RTS-TDMA essentially depends on $N_c$. As long as $N_c>N_I$, the throughput performance of our scheme is better than that of CSA. In the following, we investigate the relationship between $T_F$, $M_{max}$ and $N_c/N_I$ .\par

In Section II, a frame consists of three phases; accordingly, we have
\begin{equation}\label{formula1}
N_c\tau_c+T_f+N_t\tau_t=N_{I}\tau_{I}=T_{F},\\
\end{equation}
where $T_f$ denotes the duration of a BFP. A time slot allocation field in each feedback packet from an RSU contains only the ID of the vehicle assigned to the time slot. A feedback packet contains $N_t$ time slot allocation fields. Therefore, we assume

\begin{equation}\label{formula2}
T_f=N_t \tau_c.\\
\end{equation}
By replacing (\ref{formula2}) in (\ref{formula1}), we have

\begin{equation}\label{formula3}
\frac{N_c}{N_I}=\frac{\tau_t}{\tau_c}-(\frac{\tau_t}{\tau_c}+1)\frac{N_t\tau_I}{T_F}.\\
\end{equation}

We consider the case where $T_F$ is held constant. In RTS-TDMA, to allow all vehicles in an SZ to transmit packets together within a frame, we set $N_t=M_{max}$. Then, we have
\begin{equation}\label{formula4}
\begin{aligned}
 \frac{N_c}{N_I}=\frac{\tau_t}{\tau_c}-(\frac{\tau_t}{\tau_c}+1)\frac{M_{max}\tau_I}{T_F}.\\
\end{aligned}
\end{equation}
There is a negative correlation between $N_c/N_I$ and $M_{max}$. We can derive the critical maximum number of vehicles in an SZ, denoted by $M^{\ast}_{max}$, meaning that if $M_{max}=M^{\ast}_{max}$, then $N_c=N_I$. We have
\begin{equation}\label{formula5}
\begin{aligned}
 M_{max}^{\ast}=\frac{T_F}{\tau_I}\frac{\tau_t-\tau_c}{\tau_t+\tau_c}\\
\end{aligned}
\end{equation}
In this case, when $M_{max}< M_{max}^{\ast}$, we have $N_c>N_I$, indicating that RTS-TDMA can perform better than CSA in terms of throughput for a constant $T_F$. However, when $M_{max} \geq M^{\ast}_{max}$, we have $M_{max}\to N_I$, indicating that both schemes suffer from significantly poor throughput performance.

We consider the case where $M_{max}$ is held constant. By replacing $N_t$ in (3) with $M_{max}$, there is a positive correlation between $N_c/N_I$ and $T_F$. We denote the critical duration of a frame by $T^{\ast}_F$, which means that if $T_F=T^{\ast}_F$, then $N_c=N_I$. Note that we have
\begin{equation}\label{formula6}
\begin{aligned}
 T^{\ast}_F=M_{max}\tau_I\frac{\tau_t+\tau_c}{\tau_t-\tau_c}> M_{max}\tau_I.\\
\end{aligned}
\end{equation}
Denoting $N^{\ast}_I=T^{\ast}_F/\tau_I$, we then have
\begin{equation}\label{formula7}
\begin{aligned}
\frac{M_{max}}{N^{\ast}_I}=\frac{\tau_t-\tau_c}{\tau_t+\tau_c}.\\
\end{aligned}
\end{equation}
In general, we have $\tau_t \gg \tau_c$, resulting in $M_{max} \to N^{\ast}_I$. In this case, if $T_F \leq T^{\ast}_F$, we have $M_{max}/N_I \geq M_{max}/N^{\ast}_I$, indicating that both schemes suffer significantly poor throughput performance under the congested  scenario ($M$ is close to $M_{max} \to N^{\ast}_I$).
When $T_F > T^{\ast}_F$, we have $N_c/N_I>1$, indicating that RTS-TDMA can perform better than CSA in terms of throughput for any $M = 2, 3, \dots, M_{max}$.

\subsection{Numerical simulation}
\begin{figure}
\centering
\includegraphics[width=6.88cm,height=6.1cm]{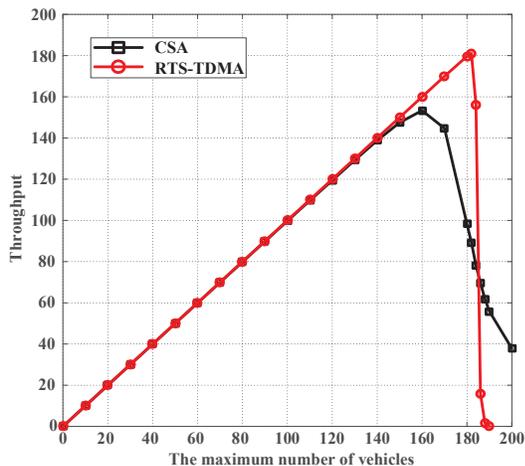}
\caption{Throughput versus the maximum number of vehicles for RTS-TDMA and CSA.}
\label{fig: simulation1}
\end{figure}

\textit{Example 1:}
In this example, we set $T_F=100$ ms as a constant and investigate the throughput performance of RTS-TDMA and CSA by varying $M_{max}$. We maintain $N_t=M=M_{max}$ during the simulation. In both schemes, to limit the number of pointers, the maximum repetition rate is fixed at 8. The optimal probability distribution with the maximum repetition rate is $\Lambda(x)=0.5x^2+0.28x^3+0.22x^8$\cite{Liva2011IRSA}. The transmission rate is 6 Mb/s\cite{IEEE802.11p}. Assuming that the total safety packet size is 375 bytes\cite{sae2009}, the duration of a time slot used to send a safety packet is 0.5 ms ($\tau_I=\tau_t=0.5$ ms), and $N_I=T_F/\tau_I=200$ in CSA. In RTS-TDMA, because only the ID of a vehicle and a pointer are transmitted, the number of bits for a vehicle ID and a pointer are 1 byte and 7 bytes, respectively. By adding guard periods and considering the physical layer overhead, we assume that the number of bits for a request packet is equal to 15 bytes. The duration of a mini-slot for a request packet transmission is 0.02 ms ($\tau_c=0.02$ ms). The simulation results are presented in Fig. \ref{fig: simulation1}. According to (\ref{formula5}), we have $M^{\ast}_{max}=185$. Therefore, for $M_{max}<185$, RTS-TDMA can outperform CSA according to the theoretical analysis in Section IV-A, as verified by the simulation results in Fig. \ref{fig: simulation1}. Note that for $M_{max}>192$, we have $N_c<0$, in which case RTS-TDMA is not applicable.

\begin{figure}
\centering
\includegraphics[width=6.83cm,height=6.1cm]{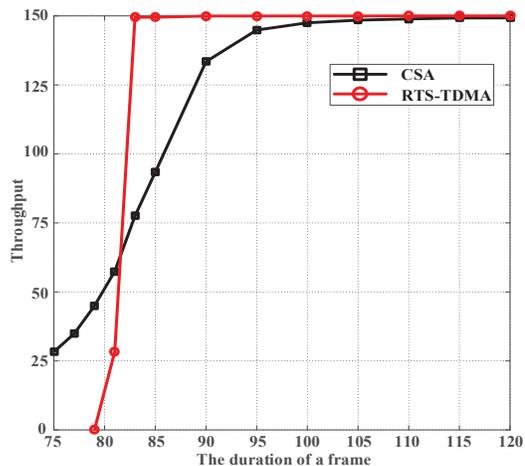}
\caption{Throughput versus the duration of a frame for RTS-TDMA and CSA.}
\label{fig: simulation2}
\end{figure}

\textit{Example 2:} In this example, we compare the performances of RTS-TDMA and CSA by setting $M_{max} = 150$ and using different durations of a frame $T_F \in \{75, \ldots, 120\}$ in milliseconds.
During the simulation, we fix $M = M_{max}$ and use the same distribution $\Lambda(x)$ as in Example 1 for selecting code lengths. We use the same $\tau_I$, $\tau_t$ and $\tau_c$ as in Example 1. The simulation results are shown in Fig. \ref{fig: simulation2}. According to (\ref{formula6}), we have $T^{\ast}_F=81.25$ ms. Thus, when $T_F>81.25$, we have $N_c>N_I$, indicating that RTS-TDMA can perform better than CSA in terms of throughput, as verified by the simulation results in Fig. \ref{fig: simulation2}. Note that for $T_F<78$ ms, we have $N_c<0$, in which case RTS-TDMA is not applicable.

\section{Conclusion}
This paper proposed RTS-TDMA, a request-transmission split TDMA scheme for vehicular networks. We divided a frame into three phases, i.e., a contention access phase, a broadcast feedback phase, and a contention-free transmission phase. To improve the reliability of the request, we repeated the transmission of each request packet, where the repetition rate was selected according to a given probability distribution. To address request collisions, we introduced the SIC technique on the RSU side. To improve the transmission efficiency, we shortened the size of the request packets, such that each request packet contained only the vehicle ID to reduce the request time percentage. The throughput of RTS-TDMA outperformed that of CSA, as verified by theoretical analysis and numerical simulations.



%
%

\end{document}